\documentclass[runningheads]{llncs}
\usepackage{todonotes}
\usepackage{graphicx}
\usepackage[caption=false]{subfig}
\usepackage{pdflscape}
\usepackage[british]{babel}
\usepackage{lscape}
\usepackage{tabularx} 

\usepackage{multirow}
\usepackage{booktabs}

\usepackage{array}
\newcolumntype{L}[1]{>{\raggedright\let\newline\\\arraybackslash\hspace{0pt}}m{#1}}
\newcolumntype{C}[1]{>{\centering\let\newline\\\arraybackslash\hspace{0pt}}m{#1}}
\newcolumntype{R}[1]{>{\raggedleft\let\newline\\\arraybackslash\hspace{0pt}}m{#1}}

\usepackage{hyperref}
\hypersetup{ colorlinks=false,       
    linkcolor=blue,         
    citecolor=blue,     
    filecolor=blue,    
    urlcolor=blue     
}
\usepackage{csquotes}

\begin{document}

\title{Compliance Requirements in Large-Scale Software Development: An Industrial Case Study}
%
\titlerunning{Compliance Requirements in Large-Scale Software Development}
%
\author{Muhammad Usman\inst{1} \and
Michael Felderer\inst{1,2} \and
Michael Unterkalmsteiner\inst{1} \and \\
Eriks Klotins\inst{1} \and
Daniel Mendez\inst{1,3} \and
Emil Alegroth\inst{1}}
\authorrunning{Usman et al.}
%
\institute{Blekinge Institute of Technology, Karlskrona, Sweden\\ \email{first.last@bth.se} \and
University of Innsbruck, Innsbruck, Austria
\and fortiss GmbH, Munich, Germany}
\maketitle              
\begin{abstract}
Regulatory compliance is a well-studied area, including research on how to \emph{model, check, analyse, enact, and verify} compliance of software. However, while the theoretical body of knowledge is vast, empirical evidence on challenges with regulatory compliance, as faced by industrial practitioners particularly in the Software Engineering domain, is still lacking. In this paper, we report on an industrial case study which aims at providing insights into common practices and challenges with checking and analysing regulatory compliance, and we discuss our insights in direct relation to the state of reported evidence.
Our study is performed at Ericsson AB, a large telecommunications company, which must comply to both locally and internationally governing regulatory entities and standards such as GDPR. The main contributions of this work are empirical evidence on challenges experienced by Ericsson that complement the existing body of knowledge on regulatory compliance.

\end{abstract}

\section{Introduction}
\label{sec:introduction}

Modern software development is driven by the problems and needs of various stakeholder groups, formulated as requirements that guide software product development. 
While a majority of these requirements are typically distilled from business stakeholders, e.g. customers or users, a group of growing importance is community stakeholders~\cite{midgley1992sacred,alexander2005taxonomy}. Community stakeholders, such as governments and other organizations, issue laws, regulations and policies, and best practices, commonly referred to as compliance requirements~\cite{akhigbe2019systematic}. 

Compliance requirements typically aim at a broad range of stakeholders and use cases, and they are, thus, purposefully expressed in general terms, omitting implementation-specific details. Elicitation and interpretation of implementation-specific details are thereby left to the expertise of affected parties, e.g. individual companies or organisations. However, while the requirements aim at providing guidance, they also pose challenges, especially in the analysis phase as stakeholders with different objectives may interpret the requirements differently~\cite{muthuri2016argumentation,ingolfo2013arguing,ghanavati2015impact,boella2013managing}.

Inconsistent interpretations may have severe consequences. 
For example, varying interpretations between development organization and regulating bodies may lead to rework, delays, financial and legal repercussions. 
This risk is exacerbated by the fact that verification of compliance is often performed late in the software development process. Consequently, any issues discovered in the compliance verification are costly to repair. 

In this study, we investigate how compliance requirements are handled at Ericsson AB\footnote{www.ericsson.com}, a large telecommunications organization. 
This study is carried out as part of a long-term research collaboration with the aim of jointly developing and improving software engineering practices in a research-centric environment. In our study at hands, we focus on understanding challenges and potential improvements in interpreting organization-wide compliance requirements for the implementation in specific use cases. 

The rest of this paper is structured as follows: Section~\ref{sec:background} presents background in regulatory compliance concepts and terms. Section~\ref{sec:related-work} provides an overview of related studies in the area. Section~\ref{sec:methodology} describes the research questions, the studied case, and outlines the research methods. Section~\ref{sec:results} presents our findings. Section~\ref{sec:discussion} discusses the findings, Section~\ref{sec:conclusion} concludes the paper.

\section{Background and Terminology}
\label{sec:background}

\emph{Regulatory compliance} is the act of ensuring adherence of an organization, process, and/or (software) product to regulations like standards, laws, guidelines, or specifications~\cite{akhigbe2019systematic}. Regulatory compliance addresses goals and mitigates risks. These goals and risks are typically linked to dependability properties~\cite{avizienis2001fundamental} like privacy, security or safety and result in regulatory artifacts like standards, laws, guidelines, or specifications.  

Compliance tasks are common activities that are carried out in order to achieve compliance with regulations. These tasks include compliance modeling, checking, analysis and enactment~\cite{akhigbe2019systematic}.

\emph{Compliance modeling} tasks address activities involved in the discovery and formalization of text extracted from regulations needing compliance. \emph{Compliance checking} tasks are activities that ensure that the formalized representations of regulations (the models) capture correctly compliance requirements. Often, activities for compliance modelling and checking are intertwined and used iteratively. \emph{Compliance analysis} tasks involve activities that provide insight into the state of compliance of an organization, a process, a (software) product, etc., as a result of the fulfilment or violation of the compliance requirements, possibly measured via models. \emph{Compliance enactment} tasks involve activities for making changes to an organization, a process, or a (software) product in order to establish or re-establish compliance with a regulation. 

The study presented in this paper covers compliance checking and analysis tasks that are performed within software and system development, which covers the core activities of software development including analysis, design, construction, and testing.

\section{Related Work}
\label{sec:related-work}

In order to collect related approaches we analyzed available secondary studies on the topic of regulatory compliance and extracted relevant primary studies on checking and analysis of regulatory requirements from them. The list of secondary studies is shown in Table~\ref{tab:secondary-studies}.

\begin{table}[htb]
\caption{Investigated secondary studies on regulatory compliance}
\label{tab:secondary-studies}
\centering
\scriptsize
\begin{tabular}{L{.45\columnwidth}L{.35\columnwidth}L{.0725\columnwidth}L{.0725\columnwidth}}
\toprule
\textbf{Title} & \textbf{Venue} & \textbf{Year} & \textbf{Ref.} \\
\midrule
A systematic literature mapping of goal and non‐goal modelling methods for legal and regulatory compliance & Requirements Engineering Journal & 2018 & \cite{akhigbe2019systematic} \\ \hline
Are we done with business process compliance: state of the art and challenges ahead & Knowledge and Information Systems Journal & 2018 & \cite{hashmi2018we} \\ \hline
Using Business Process Compliance Approaches for Compliance Management with Regard to Digitization: Evidence from a Systematic Literature Review & International Conference on Business Process Management & 2018 & \cite{sackmann2018using} \\ \hline
An extended systematic literature review on provision of evidence for safety certification & Information and Software Technology Journal & 2014 & \cite{nair2014extended} \\ \hline
A Systematic Review of Compliance Measurement Based on Goals and Indicators & Advanced Information Systems Engineering Workshops & 2011 & \cite{shamsaei2011systematic} \\  \hline
A Systematic Review of Goal-oriented Requirements Management Frameworks for Business Process Compliance & Workshop on Requirements Engineering and Law & 2011 & \cite{ghanavati2011systematic} \\
\bottomrule
\end{tabular}
\end{table}

From the secondary studies, we extracted overall 22 primary studies related to checking and analysis of regulatory and legal compliance requirements. As regulations like laws or standards are often region or even country specific, most primary studies explicitly refer to regulations from a specific region or country, i.e., USA or Canada~\cite{hassan2009governance,breaux2006algorithm,hassan2008validating,maxwell2009checking}, Latin America~\cite{da2014towards}, Australia~\cite{abdullah2010emerging}, and Europe~\cite{ingolfo2013arguing,soltana2016model}. Only laws or regulations specific to Asian countries are not covered. 

The studies also cover classical regulated domains including finance~\cite{hassan2009governance,soltana2016model}, medical~\cite{maxwell2009checking,miseldine2008supporting,ingolfo2013arguing,massey2010evaluating,breaux2006algorithm}, law~\cite{muthuri2016argumentation,ghanavati2015impact}, public administration~\cite{da2014towards}, automotive~\cite{penzenstadler2008complying}, and avionics~\cite{arthasartsri2009validation,conmy2007challenges}.

A prominent topic in the studies on regulatory requirements is the extraction or definition of models~\cite{miseldine2008supporting,soltana2016model,hassan2009governance,breaux2006algorithm} for compliance checking or monitoring. However, formal checking of compliance is resource-intensive and therefore its complexity is investigated~\cite{governatori2008detecting,tosatto2014business}. As regulatory requirements intentionally or unintentionally differ in scope, their interpretation is an important topic covered in several publications~\cite{muthuri2016argumentation,ingolfo2013arguing,ghanavati2015impact,boella2013managing}. 

Core regulated properties covered are related to security~\cite{massey2010evaluating}, privacy~\cite{massey2010evaluating,hassan2008validating} and safety~\cite{penzenstadler2008complying,arthasartsri2009validation,hu2009experience}.

Most papers are either validated by examples or evaluated based on case studies. Only a few surveys on regulatory compliance are available. Abdullah et al.~\cite{abdullah2010emerging} present a survey on emerging challenges in managing regulatory compliance. The report~\cite{pwc2015survey} presents a survey on the state of regulatory compliance in practice.

50\% of the 22 papers directly address software development and its artifacts. Most papers cover compliance modelling followed by compliance checking, analysis and enactment. Challenges explicitly mentioned are the complexity and ambiguity of regulations by nature~\cite{massey2010evaluating}, documentation and modeling of the relevant regulatory constraints and their derivations, establishing and keeping traceability of artifacts as well as change management and evolution for keeping the documents, history, and references up to date~\cite{penzenstadler2008complying,soltana2016model}, and simple, easy-to-use and aligned modeling and reasoning approaches~\cite{soltana2016model,hu2009experience}.

Via snowball sampling from the related primary studies, we found another very closely related paper to our work by Nekvi et al.~\cite{nekvi2014impediments}. In that paper, the authors present impediments to regulatory compliance of requirements in contractual systems engineering projects based on a case study from the railway domain. 

In addition, several other papers, not covered by the secondary studies, discuss software engineering aspects of regulatory compliance. Hamou-Lhadj~\cite{hamou2015regulatory} discusses regulatory compliance and its impact on software development in general. Furthermore, a compliance support framework for global software companies is provided in~\cite{hamou2007towards}. Several authors discuss the relationship between agile software development and regulatory compliance. Mishra and Weistroffer~\cite{mishra2008issues} discuss issues with incorporating regulatory compliance into agile development.

In the medical domain, McHugh et al.~\cite{mchugh2012barriers} discusses barriers to adopting agile practices when developing medical device software. Soltana et al.~\cite{soltana2016model} refine their UML-based compliance checking approach to enable checking against the GDPR~\cite{torre2019using}. Fitzgerald and Stol~\cite{fitzgerald2017continuous} shape the term "continuous compliance" expressing that software development seeks to satisfy regulatory compliance standards on a continuous basis, rather than operating a "big-bang" approach to ensuring compliance just prior to release of the overall product. Finally, Hashmi et al.~\cite{hashmi2018we} discuss the potential of blockchain to compliance adherence in distributed software delivery.

Synthesis of this body of work shows that regulatory compliance is a well studied area.
However, whilst the theoretical foundation is well established, the amount of empirical industrial case studies~\cite{nekvi2014impediments} and surveys~\cite{abdullah2010emerging,pwc2015survey} on compliance checking and analysis is limited rendering our understanding of what challenges industry faces still weak. This understanding, however, is imperative to steer research activities in a problem-driven manner.
This leads us to conclude that there is a gap of knowledge about the regulatory compliance challenges experienced by industrial practitioners.
Thereby providing motivation for this study and support for the value of its contribution.

\section{Research Methodology}
\label{sec:methodology}

Our goal was to develop support for understanding the current processes and challenges associated with checking and analysis of compliance requirements in a large-scale software development context. To do so, we conducted an exploratory case study \cite{runeson2009guidelines}. In this section, we describe the research questions, the case, and also the data collection and analysis methods.

\subsection{Guiding Research Question}

We framed the following guiding research question to steer our study:

\begin{center}

    \emph{What are the challenges and potential improvements for checking and analysis of the compliance requirements?}

\end{center}

The research question aims at identifying the challenges faced along the process of checking and analysing the compliance requirements and also the potential improvements to address the challenges.

\subsection{Case description}

The case company is Ericsson AB, a large multinational company developing software-intensive products related to Information and Communication Technology (ICT) for service providers. Besides the business requirements of the individual products, there are some generic requirements that are applicable to the entire Ericsson portfolio including over 100 products. The compliance with these generic requirements is mandatory for all products. In this study, we refer to these requirements as compliance requirements. Compliance requirements are specified and maintained by a central unit at the organization level. The central unit also proposes and maintains the design rules and guidelines for supporting the development teams in implementing the compliance requirements.

The case under investigation is a large software product in the telecommunication domain that consists of several sub-systems. Multiple teams, which are  geographically distributed, are involved in the development of each sub-system. In the studied case product, the following organisational units work together to handle the compliance requirements:

\begin{itemize}
    \item Product management: The product management unit selects the compliance requirements that apply to the release under consideration.
    \item System management: System management, together with the product management, performs the initial analysis to identify the impacted sub-systems. In this initial analysis step, the system management tries to clarify the compliance requirements by reducing assumptions with additional information.
    \item Sub-systems' development unit: The development unit of each impacted sub-system is responsible for ensuring compliance with the selected compliance requirements, for example, by using the recommended design rules during implementation.
    \item Verification unit: In the end, the verification unit performs the verification of compliance requirements. The verification process results in a compliance report identifying the sub-systems that are compliant (fully or partially) and non-compliant with the compliance requirements.
\end{itemize}

The unit of analysis in this study is the compliance process used by multiple units and roles in the case company who specify, analyse, plan, implement and verify compliance requirements.

\subsection{Data collection and analysis}
To maximize the validity of our results, we used a combination of different research methods to collect the data.

\begin{itemize}
    \item Group discussions - We performed two open-ended group discussions involving multiple researchers and practitioners. The first one was performed at the very beginning of the study to define the scope of the study and agree on a plan. The group discussion involved two researchers (first and third author), unit manager (our main contact for the study) and a member of the verification unit of the case product. The second group discussion was performed towards the end of the study, wherein the researchers (first, second and the fourth author) presented the results and together with the practitioners, representing different units related to the compliance work, identified the future course of actions
    \item Workshop - To understand the current state-of-the-practice on the verification of the compliance requirements, we conducted a three-hour workshop involving the following roles from the case product: one product manager, one system manager, two architects from two different sub-systems, one product owner and one deployment lead. The verification unit was not represented in the workshop, instead we interviewed them separately to ensure that all the participants feel free to express their opinions. The first author lead the workshop with the support of the third and fourth author. The third and fourth author independently took notes to capture the discussions. During the workshop, we used an instrument (see Section \ref{sub:instrument} and Figure \ref{fig:example_instrument} for more details about the instrument) to collect data. 
    
    We conducted the workshop according to the following plan:
    \begin{itemize}
        \item Introduction (15 mins): To introduce the participants to each other and to describe the purpose of the workshop
        \item Description of the data collection instrument and the remaining workshop activities (15 mins)
        \item Data collection using the workshop instrument (30 mins)
        \item Break (15 mins)
        \item Open discussion and presentations of the collected data (60 mins)
        \item Prioritize the ideas to work in the future (30 mins)
        \item Summary (15 mins)
    \end{itemize}
     \item Interviews - The workshop did not have the representation from the central unit that specifies the compliance requirements nor the verification unit that performs the testing. To cover these two units, we performed two semi-structured interviews that lasted for about an hour each: one with the information owner who is responsible for writing and maintaining the compliance requirements, and second with a test manager of the unit responsible for verifying the compliance requirements.
    \item Process documentation - We also received and reviewed the official description of the processes related to the compliance work in the case product.
\end{itemize}

We analyzed the qualitative data about challenges collected during the workshop and the interviews by applying coding \cite{miles1994qualitative}. We used the knowledge obtained from the related works and during the initial group discussions to assign interpretative codes \cite{miles1994qualitative} to group the challenges into different categories (e.g., process related challenge). The first author assigned these interpretive codes to categorize the challenges. The proposed categories were presented to the other authors and also to the workshop participants.

\subsection{Workshop instrument}
\label{sub:instrument}

In the product under investigation, the compliance work includes various inter-dependent activities connected through their inputs and outputs. The practitioners involved in the initial group discussion highlighted the importance of alignment between different activities. To systematically collect the data about different compliance activities involving multiple units, we designed a workshop instrument (see Figure \ref{fig:example_instrument}). We shared the workshop instrument with the unit manager of the verification team one week before the workshop. We requested the unit manager to review the instrument and provide us their feedback. The unit manager did not suggest any changes in the instrument. The workshop instrument consists of the following parts:

\begin{figure*}[ht!]
    \centering
    \includegraphics[scale=0.45]{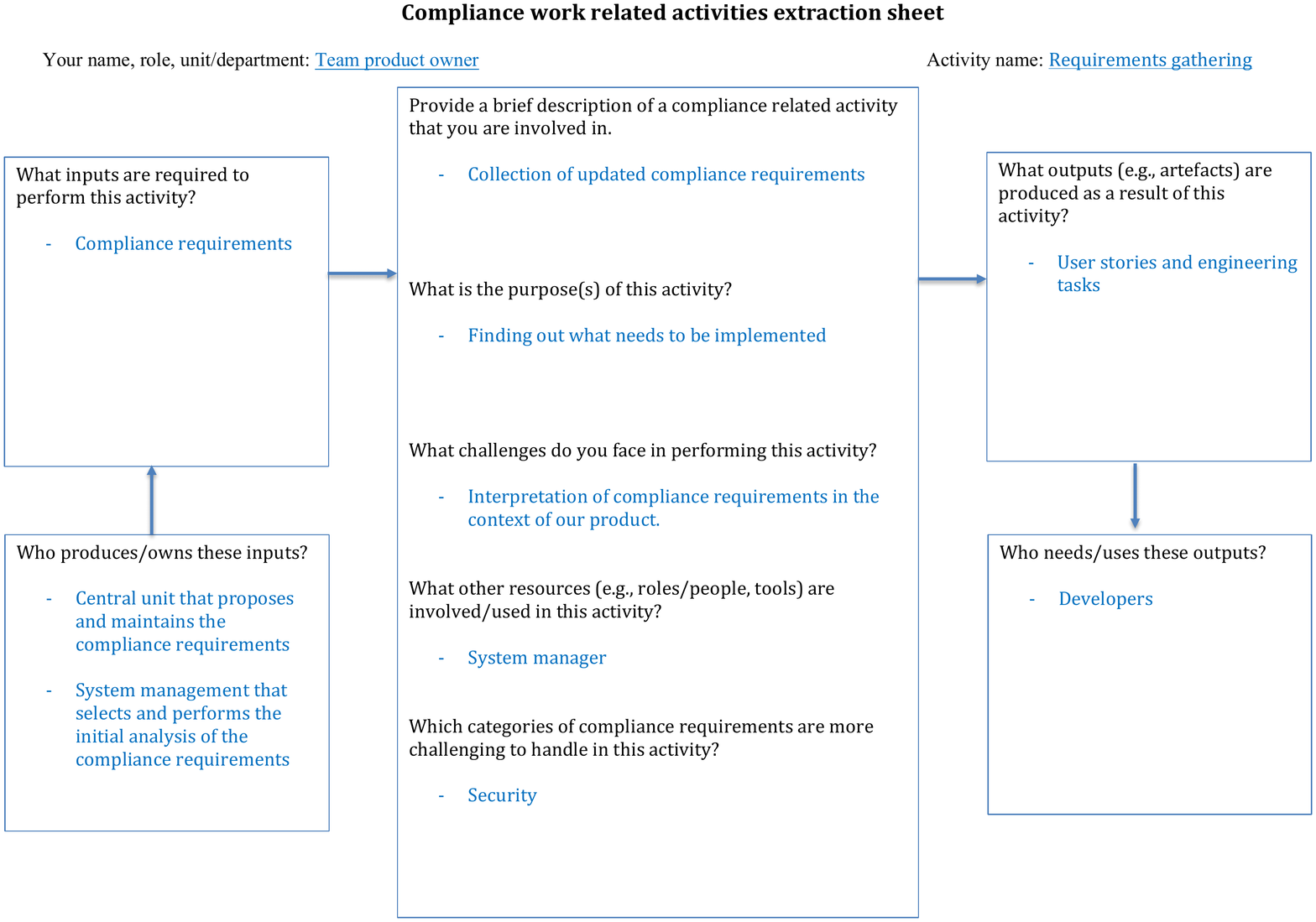}
    \caption{Instrument used in the case study (one instance filled out by one of our workshop participant).}
    \label{fig:example_instrument}
\end{figure*}

\begin{itemize}
 \item Activity: Each role relevant to the compliance work is responsible for one or more activities. In this part of the instrument, the workshop participants were asked to describe one compliance work-related activity in which they are involved. The instrument also captures the challenges that the participants face while performing the activity. The activity may involve other resources (e.g., other roles or tools). Therefore, the instrument also has a question about other resources used during the activity, if any. Lastly, some categories of compliance requirements (e.g., security related compliance requirements) are more challenging to handle. The instrument covers the categories of the compliance requirements as well to identify the one that are more challenging to handle.
\item Inputs and their owners: Compliance activities depend on certain inputs that are owned or produced by other roles. For example, compliance requirements are described and maintained by a central unit. To understand the alignment and expectations of different roles from each other, it is important to capture if different roles have a shared understanding with regards to inputs and their ownership. 
\item Outputs and their users: Like inputs, the instrument also captures the outputs produced by the described activity and roles that need that output.
\end{itemize}

\section{Results}
\label{sec:results}

In the following, we report on the study results. Note that our discussion of those results in relation to the existing body of knowledge is provided in subsequent sections.

\subsection {Overview}

In total, seven practitioners from the case product and one from the central unit participated in the study: six in the workshop and two during the interviews. During the workshop and the interviews, the participants highlighted several challenges related to the compliance requirements. The identified challenges are related to the compliance analysis and checking tasks (see Section \ref{sec:background} for details on the four types of compliance tasks). The modeling task only came under discussion during the interview with the participant from the central unit that specifies and maintains the compliance requirements. Since the workshop participants do not specify compliance requirements, the focus during the workshop was on interpretation of the requirements, rather than their modeling. Furthermore, the enactment task did not come under any discussion during the interviews or the workshop. The findings of the current study may result in the initiation of some enactment related tasks. Besides challenges, the participants also provided several ideas to further improve the compliance work.

We analyzed the identified challenges and grouped them into three categories: requirements specification related challenges, process related challenges and resource related challenges. We now discuss these challenges and associated improvements suggested by the study participants.

\subsection{Requirements specification related challenges}
In this category, we included the challenges related to the way the compliance requirements are specified. Compliance requirements are not written for a specific product; instead, they are specified at an abstract level as general requirements that are applicable for all products in the case organization.  All participants highlighted the challenge of interpreting the compliance requirements in the context of their product. They view the interpretation of the compliance requirements as the most challenging aspect of the compliance work. Furthermore, at times different units interpret the same compliance requirement differently, which results in disagreements about the expectations. System management is already complimenting the compliance requirements with some additional information to explain them better. The workshop participants suggested to take the additional explanation idea forward and develop a shared interpretation of the compliance requirements in the context of the product.

Some compliance requirements conflict with other requirements (e.g., security and usability), and it becomes challenging to perform the trade-off analysis. In the case product, the business requirements are handled as use cases, while compliance requirements are handled separately. The development teams find it challenging to focus on both business and compliance requirements. Lastly, the central unit that specifies and maintains the compliance requirements, also proposes the design rules to help in the implementation of the compliance requirements. However, in some cases, the link between compliance requirements and the corresponding design rules is not easy to follow. Table \ref {tab:challenge_description} lists the requirements specification related challenges.

\fbox{\begin{minipage}{11cm}
		\footnotesize
		Challenges: \textit{\enquote{...Interpreting the compliance requirements, how the general high-level requirements apply to the environment that the product exists within...}
		}\\
		\hspace*{\fill} Product Owner
		
		Challenges: \textit{\enquote{...Interpretation of the compliance requirements into our product's context...}
		}\\
		\hspace*{\fill} System Manager
		
		Challenges: \textit{\enquote{...Understanding the intent of the requirements...}
		}\\
		\hspace*{\fill} Architect
	\end{minipage}
}

\begin{table*}[htbp!]
    \centering
    \caption{Requirements specification related challenges}
    \begin{tabular}{l}
        \hline
        Challenge description \\ \hline
        - Interpretation of compliance requirements in the context of a specific product  \\
        - Differences in the understanding of the compliance requirements \\
        - Abstractness of the compliance requirements  \\
        - Trade-offs and conflicts between different compliance requirements \\
        - Missing linkage with the business use cases  \\
        - Linkage between the compliance requirements and design rules \\
        \hline
    \end{tabular}
    \label{tab:challenge_description}
\end{table*}

\subsection{Process related challenges}
In this category, we included the challenges related to the compliance process. The participants highlighted the need to further improve the alignment between different compliance activities. The participants also suggested to improve the coordination between different roles and units involved in the compliance process. In particular, the coordination between the verification unit and the development unit in each sub-system needs to further improve. The case product has several sub-systems, which do not have a consistent process of handling the compliance work. 

Furthermore, at times different categories of compliance requirements (e.g., security) are handled differently. The participants highlighted the need to have a consistent process for all compliance requirements. They suggested to further improve the documentation of the entire compliance process with the aim to improve the alignment between different units and to clarify the involved roles and their responsibilities. The participants also shared the view to introduce better requirements management tools to effectively manage and communicate the compliance requirements. Additionally, since compliance requirements and business requirements are managed separately, the participants identified the need to establish a balance between the two types of requirements. Lastly, compliance tasks are not fully automated yet, which makes it a time-consuming and effort-intensive work. Table \ref {tab:challenge_process} lists the process related challenges.

\fbox{\begin{minipage}{11cm}
		\footnotesize
		Challenges: \textit{\enquote{...Different processes used in different organizations resulting in bumpy compliance project requiring extra alignment and project management...}
		}\\
		\hspace*{\fill} Deployment Lead
		
		Challenges: \textit{\enquote{...Balance between functional and compliance requirements development...}
		}\\
		\hspace*{\fill} Product Manager

	\end{minipage}
}

\begin{table*}[htbp!]
    \centering
    \caption{Process related challenges}
    \begin{tabular}{ l }
        \hline
        Challenge description \\ \hline
        - Coordination and alignment of the compliance tasks between sub-systems' teams\\
        - Compliance requirements not communicated properly  \\
        - Different compliance requirements (e.g., security) managed differently  \\
        - Missing dedicated process at the sub-system level  \\
        - Lack of coordination between verification and development teams \\
        - Change management of compliance requirements \\
        - Establishing a balance between compliance and business requirements\\
        - Prioritising the right compliance requirements \\
        - Lack of automation\\
        \hline
    \end{tabular}
    \label{tab:challenge_process}
\end{table*}

\subsection{Resource related challenges}
In this category, we included challenges that are related to the resources (people, tools) required for handling the compliance requirements. All participants highlighted the lack of available resources and time for handling the compliance requirements. Business requirements that are handled separately consume most of the capacity of the development teams. The teams find it hard to allocate enough resources to handle the compliance requirements. Furthermore, compliance work requires much coordination with other units, which means additional time and resources. The participants also highlighted that the developers are relatively less aware of the compliance requirements and associated design rules as compared to the business requirements. The participants suggested to increase and expand the existing training programs and initiatives to improve the awareness and knowledge about compliance requirements and design rules. Lastly, participants also pointed out that there is a need to introduce better tools to manage the compliance requirements. Table \ref {tab:challenge_resource} lists the resource related challenges.

\begin{table*}[htbp!]
    \centering
    \caption{Resource related challenges}
    \begin{tabular}{ l }
        \hline
        Challenge description \\ \hline
        - Lack of dedicated resources and time to handle compliance requirements \\
        - Lack of awareness among developers about compliance requirements \\
        - Lack of awareness among developers about design rules  \\
        - Tools used to manage compliance requirements are not appropriate \\ 
        \hline
    \end{tabular}
    \label{tab:challenge_resource}
\end{table*}

\section{Discussion}
\label{sec:discussion}

While research on supporting modelling, checking, analysing and enacting
compliance tasks is abound, as witnessed by the secondary studies discussed in Section~\ref{sec:related-work}, studies that
investigate compliance challenges in practice are
rare~\cite{nekvi2014impediments}. Furthermore, most of those studies base their
findings on educated opinion or experience~\cite{nekvi2014impediments}. In this
paper, we contribute to the body of knowledge of regulatory compliance
challenges, initially compiled by Nekvi and
Madhavji~\cite{nekvi2014impediments}, by adding the case study reported in
Section~\ref{sec:results} and the studies listed in
Table~\ref{tab:challenge_studies}, which are discussed next. We further provide a brief discussion on fruitful experiences when using this rather unconventional instrumentation before  concluding with a discussion of the threats to validity and the mitigation measures.

\subsection{Results in Relation to Existing Evidence}

\begin{table}
  \centering
  \caption{Studies on challenges in compliance requirements}
  \begin{tabular}{lll}
    \hline
    Paper & Context & Type of study \\ \hline
    Abdullah et al.~\cite{abdullah2010emerging} & Compliance management & Case study \\
    Conmy and Paige~\cite{conmy2007challenges} & Safety standards (avionics) & Educated opinion \\
    Boella et al.~\cite{boella2013managing} & Business processes & Educated opinion \\
    Ghanavati et al.~\cite{ghanavati2015impact} & Business processes & Experience \\
    Nekvi and Madhavji~\cite{nekvi2014impediments} & Railway regulations & Case study
\\    \hline
  \end{tabular}
  \label{tab:challenge_studies}
\end{table}

Abdullah et al.~\cite{abdullah2010emerging} grouped compliance management
challenges into customer, regulation and solution related factors. Common
challenges to our study are the lack of connecting compliance to business
objectives, the lack of communicating a common understanding of compliance
continuously to employees, inconsistencies in applying regulations, the
lack of compliance practices applied throughout an organization, and the lack of
tool support for compliance management and monitoring.

Conmy and Paige~\cite{conmy2007challenges} found that the reuse of models in
model-driven architecture (MDA) development processes is challenging when there is a
need to certify artefacts. With simple and reusable models, the certification
process becomes more burdensome as information is spread between artefacts. On the
other hand, if information is duplicated and/or organized such that
certification becomes easier, the model artefacts become harder to maintain and
reuse, reducing the benefits of MDA.

Both Boella et al.~\cite{boella2013managing} and Ghanavati et
al.~\cite{ghanavati2015impact} report on the difficulty of interpreting
regulations. The common identified challenges are (1) the need to interpret generic
regulations such that they are implementable in specific situations; (2) the
communication between stakeholders with different background and viewing
compliance issues from their perspective and language; and (3) compliance
dynamics, referring to changes in regulations or products that are regulated,
and the need to adapt efficiently and effectively to such changes.

Finally, Nekvi and Madhavji~\cite{nekvi2014impediments} found the following
common challenges: (1) identifying and accessing relevant set of regulatory
documents; (2) the number, size and complexity (in terms of cross-references) of
regulations; and (3) cross-cutting concerns of regulations that span different
subsystems and require communication between development groups.

Looking at the studies that investigate regulatory requirements, we
observe that few developed a dedicated data collection instrument for
systematically identifying challenges. Only Nekvi and
Madhavji~\cite{nekvi2014impediments} and Abdullah et
al.~\cite{abdullah2010emerging} conducted case studies and reported detailed data
collection and analysis procedures while the majority of the research relies on
experience and educated opinion reports.

Looking at the results of our case study, we find support for all the challenges
reported at Ericsson AB. Hence, we conclude that, while the business domains and
contexts may differ, there exists a core of common challenges to regulatory
requirements that are in need for more solution-oriented research.

\subsection{Discussion of Research Methodology}

In our workshops, we used a specifically designed instrument to connect different compliance activities using their inputs, outputs and owners/users and to identify the state of compliance practice through which the compliance requirements are analyzed, implemented, and verified. Furthermore, it also aims at capturing the challenges involved in the entire compliance work. The workshop participants were asked to fill one instrument per activity. In total, six company participants filled 11 such instruments (hand-written on A3 paper, see one transcribed example in Figure \ref{fig:example_instrument}). Next, the participants were asked to present their instruments during one hour open discussion session. 

Overall, we noticed that this discussion facilitated the participants to understand the perspectives of other roles, and also to develop a shared understanding of the compliance work. In particular, the discussion on the challenges faced by each role helped the other roles to realize the importance of better coordination and alignment.

The workshop instrument therefore helped us to not only collect the relevant data efficiently, but it also supported us in structuring the discussion session around the main themes involving compliance activities and their purpose, challenges, categories of the compliance requirements that are more difficult to handle, and the inputs and outputs of the activities. The participants themselves explicitly appreciated the idea of using such a structured instrument during the workshop which is also the reason why we cordially invite interested researchers and practitioners to employ it (or variants of it) themselves report in their respective environments. In our personal view and experience from using it, the workshop instrument provides a good support in understanding the state of compliance practice in large-scale software projects wherein multiple units need to work together to handle the compliance requirements. 

\subsection{Threats to validity}

As any other empirical study, ours has faced several threats to validity. Here, we report on the most prominent ones related the sampling strategy, subjectivity, and to generalisation.

\emph{(Convenience) Sampling.} Our choice of industry partner emerges from the pre-existing relationship with them and can be seen as opportunistic. However, we argue that the company is well established and a representative candidate for the research area under investigation. Further, we were able to draw from the insights and experiences of a broad spectrum of stakeholders and, thus, are confident about the completeness in the views and expressed opinions of the involved parties to be able to draw the conclusions we have drawn.

\emph{Subjectivity.} Subjectivity always plays a vital role in qualitative studies and ranges from several forms of bias of the participants over (mis-)interpreting others' opinions and statements to potential inaccuracies in the coding. While we were particularly interested in gathering subjective experiences and opinions, we were able to mitigate threats emerging from bias and mis-interpretations potentially distorting our results such as by the trustful long-enduring relationship with the partner, by re-assuring the participants transparently about the scope of the study and how we handle the data over them always having the possibility of re-evaluating our results along the final presentation. Especially the latter strengthens our confidence in the accuracy of our coding and in the correctness of how we have captured and interpreted the results. 

\section{Conclusion}
\label{sec:conclusion}

In this paper, we investigated the state of practice in regulatory compliance with an industrial case study. Our  primary contribution are the insights gathered on the challenges and a mapping towards existing literature, of what challenges with regulatory compliance can be observed in industrial practice in Ericsson.

To this end, we have further developed a particular workshop instrument. This instrument was effective to identify both the regulatory compliance process used by Ericsson but also gaps and challenges within this process.
As such, we perceive this instrument as valuable both for researchers studying the area of regulatory compliance as well as practitioners that wish to evaluate their own compliance process and practices.

\textbf{Acknowledgments.} We would like to acknowledge that this work was supported by the Knowledge Foundation through the projects SERT – Software Engineering ReThought and OSIR (reference number 20190081) at Blekinge Institute of Technology, Sweden.

%
%
%
\bibliographystyle{splncs04}
\bibliography{references}

\end{document}